\documentclass[12pt,a4]{article}
\usepackage[utf8]{inputenc}
\usepackage{amsmath}
\usepackage{url}
\usepackage{bbold}
\usepackage[english]{babel}
\usepackage{graphicx}
\usepackage{siunitx}
\usepackage{xcolor}
\usepackage{amsfonts}

\newcommand{\rms}{\rho} 
\newcommand{\srms}{\varrho} 
\newcommand{\mgt}{m} 
\newcommand{\mlr}{\hat m} 
\newcommand{\miw}{\tilde m} 

\title{How accurate can combined measurements be~-- experiment, simulation, and theory}
\author{B. Mirbach\thanks{\quad German Research Center for Artificial Intelligence, Department of Augmented Vision, Trippstadter Str. 122, 67663 Kaiserslautern, Germany; bruno.mirbach@dfki.de},
M. Boguslawski\thanks{\quad University of Münster (alma mater), 48149 Münster, Germany; martin.boguslawski@mailbox.org}}
\date{May 2022}
\begin{document}
\maketitle
\begin{abstract}
In this paper we investigate the question of how much combined measurements can increase the accuracy of additive quantities.
Therefore, we consider a set of measurements from a selection of all possible combinations of the $n$ labeled masses and then estimate the individual weights of the $n$ masses by a linear regression approach.

We present experimental results which motivate comprehensive simulation campaigns. 
These simulations provide valid statistical statements and reliable forecasts of the experimental results.
A profound analytical treatment in turn supports these simulation outcomes with excellent consistency.
One important achievement therein is a general analytical expression for the estimate's error, not only limited to the two particular weighing schemes presented.

It turns out that combined measurements allow to estimate the weight of mass elements with an accuracy that under-runs by orders of magnitude the resolution of the scale used.
As the error depends on the amount of measurements, one gains higher accuracy with increasing effort.

In a broader sense, our work wants to promote the method and give inspirations to applications in various metrological fields beyond high-precision mass determination. 
Moreover, the novel simulations and analytic formulas enable the design of optimal experiments.

\end{abstract}
\newpage
\tableofcontents
\newpage

\section{Introduction}
It is said that coffee has an inspiring effect.
This true piece of wisdom can only be acknowledged by the authors; as the central question of this paper arose from following a strict recipe to prepare a good and reproducible espresso: 
"Can a usual (kitchen) scale weigh a coffee bean more precise than the scale's reading precision would allow?".
In a more provocative version: "Can a single rice corn be weighed with a truck scale?". 
These and similar questions open out to the central hypothesis of this paper: 
Considering numerous combinations out of a set of labeled mass elements leads to higher precision of the weight of a single entity than its individual weighing alone.
\\
The result of a scale, being analogue or digital, is obviously rounded towards its reading precision.
There are, however, particular cases where weighing a set of masses can give more accurate information about the individual mass.
Imagine a set where the variance of the weights is far below the precision of the scale.
The total weight of all elements divided by the number of elements will allow to estimate the single masses more precise than the precision limit. 
However, if the variance is somewhere above the scale precision, this approach fails as nothing is gained by determining average weights.

Early approaches of mass standard estimation supported by linear regression were published in the textbook "Praktische Physik" (German for practical physics) by F. Kohlrausch in 1870, meanwhile established as a common build-up method in the field of mass metrology.
The relevant chapter therein about mass determination is based on a linear equation system, 
and was part of the book until its last edition in 1996
\cite{Kohlrausch:1996}. 
In the chapter 1.1.4.3 "Masse\-skala" (German for mass scale) of the 1996 edition, a weighing scheme is suggested to create a mass scale over several orders of magnitude, starting with only one known mass element of small relative uncertainty (reference mass) among a set of unknown masses.
The idea behind is to build up a chain of measurements with each two comparable total masses balanced against each other, and beginning with the reference.
In this sense, a linear equation system is set up whose solution is minimized by linear regression approaches.
Such a procedure allows to precisely estimate the unknown masses within small uncertainty.

A similar experimental procedure is presented in several publications (e.g. \cite{Glaeser_2009, Borys:2012:Springer}) around the Physi\-kalisch-Technische Bundesanstalt (PTB) located in Braunschweig, Germany. 
On the occasion of the 125th anniversary of the institute, 
the PTB put online a digital open-access copy of the \textit{Kohlrausch} and refers to its close historic connection to this important textbook since firstly published. 
Rather than insufficiently giving a wrap-up of the history of mass determination and metrology, we want to refer to the all-encompassing reviews and details within the mentioned publications and numerous references therein.

The method of least square fitting goes back more than 200 years to Legendre and Gau{\ss}, who used it to fit orbits to astronomic observations (see e.g. \cite{wikipedia:Regression,Freedman:2009} and references therein). In the late $19^{\text{th}}$ century, the method was used to fit biological and social science models. 
In recent times, least square regression has become an important tool in machine learning \cite{bishop:2006:PRML}.
The reason for the large success of least square regression in all fields of science is expressed in the \textit{Gau\ss-Markov} theorem which states that minimizing the square error of a measurement yields the best linear unbiased estimate in case the errors are uncorrelated, have equal variances, and an expectation value of zero.
If the measurements have non-equal error variance, one may normalize the errors by their known variance to obtain an optimal regression result (see e.g. \cite{Freedman:2009}).
A recent general examination about this so-called weighted measurement method was presented in \cite{Wiora2016, Ho2020} with a focus on weighted estimations and numerous applications beyond mass metrology.
In case of error correlations, the more generic \textit{Gau\ss-Markov} approach is suggested which minimizes the error covariance \cite{Kruskal_1968}. 
In \cite{Bich_1990} this method has been used to perform estimates with mass restraints of known uncertainties incorporated, being a classical problem in mass metrology \cite{Borys:2012:Springer}.

A combinatorial method is discussed in \cite{White2008} as a general calibration technique for indicating instruments.
Therein, a variety of applications in the field of mass metrology, optical detectors, and resistance bridges is presented.
In 2012, Siuda and Grabowski \cite{SIUDA20121165} came up with the idea of using the combined measurements approach to improve the accuracy of measurements of additive quantities, e.g. without the need of any mass standards of known uncertainty.
Their idea supposedly came up rather independent from previous work done in the field of mass metrology, as only a few references and no state of the art are given therein.
The focus is fully put to the feasibility of the idea with brief theoretical estimations, simulations, and measurements that are limited to low $n$ (e.g. $n = 5$).
A following "comment on..." paper \cite{WIORA20132259} picked up the discussion about how strong the additive quantities correlate, giving a comparison between outcomes of correlated and uncorrelated considerations.

The combined measurements approach presented in \cite{SIUDA20121165} is rather similar to our initial idea.
In this publication, however, we will go into more detail with deeper insights gained by theoretical considerations that are supported by simulations and comprehensive experimental verification.
Thus, a more general expression for the uncertainty of estimated masses is established, depending on the measurement design comprising the weighing scheme and all measurement parameters.
We consequently see in our work a valuable contribution and, moreover, want to help this approach to gain general visibility beyond an important, however rather specialized scientific sector. 
Our work wants to motivate interested and exploratory people to develop further inspiring applications in diverse fields of metrology, and to provide guidance for designing optimal combined measurement schemes as required. 

The further structure of this paper is as follows: In section \ref{sec:concept}, we discuss the basic concept of the combined measurements method introducing the nomenclature, as well as fundamental equations and the approach of error minimization via linear regression estimation.
Section \ref{sec:Experiment} presents the set-up and results for an experimental proof of concept. 
We therein demonstrate that it is possible to accurately determine the weights of small stones with a common kitchen scale that has only a rough reading precision of the order of the stone weights. 
Therefore, we put forward two particular weighing schemes of different levels of experimental effort and estimation accuracy.
In section \ref{sec:Theory}, we first briefly recapitulate the mathematical basics of linear regression by least square fitting. 
We then derive a generic expression for the covariance matrix of the estimated parameters, which reveals the dependence of the regression accuracy on the number of parameters (i.e.~the set size of masses) and on the number of experiments.
Applying these findings to the weighing scheme cases presented in the experimental part, we give particular analytic expressions for the regression errors.
Corresponding simulation results are given in section \ref{sec:Simulations}, again picking up the two weighing schemes with a comparison of experiment, theory, and comprehensive simulations.

\section{Concept -- the linear scale model}
\label{sec:concept}
We assume a mass scale to measure a weight $w$ according to the linear model
\begin{eqnarray}
\label{eq:linModel}
y(w) = \mgt_0 + w + \varepsilon(w)
\end{eqnarray}
with $\mgt_0$ being a constant offset and $\varepsilon$ an error with variance $\sigma^2$ and mean zero. 
Further, we consider in the following the weight $w$ being a combination of $n$ unknown
mass elements $\{\mgt_i, i=1 \dots n\}$, i.e. 
\begin{equation}
\label{eq:sumWeight}
w = \sum \limits_{i=1}^n x_i  \mgt_i .
\end{equation}
Thereby, the coefficients $x_i \in \{0,1\}$ indicate if the corresponding mass $\mgt_i$ contributes to the weight or not. 
The model (\ref{eq:linModel}) thus becomes
\begin{equation}
\label{eq:linEqSys}
y = \mathbf{x}\mathbf{\mgt} + \varepsilon
\end{equation}
with $\mathbf{\mgt} = (\mgt_0,\mgt_1,\ldots,\mgt_n)^T$ and $\mathbf{x} = (1,x_1,\ldots,x_n)$. Here the column vector $\mathbf{\mgt}$ of the masses has been extended by one element for the offset $\mgt_0$ and the row vector $\mathbf{x}$ of coefficients by a corresponding constant element $x_0\!=\!1$.\\ 

Performing $N\ge n+1$ measurements $\{y_j, j=1,\dots,N\}$ from different combinations $\{\mathbf{x}_j, j=1,\dots,N\}$ of the $n$ unknown masses elements $\mgt_i$ allows to determine these masses as well as the offset $\mgt_0$ via linear regression.
Therefore one minimizes the square error
\begin{align}
\label{eq:SEdef1}
E = \sum \limits_{j=1}^N \varepsilon_j^2 &= \sum \limits_{j=1}^N |y_j - \mathbf{x}_j \mathbf{\mgt}|^2 .
\end{align}

The common way to find the minimum of the square error is to solve a linear equation derived from (\ref{eq:SEdef1}). 
In section \ref{sec:Theory}, we discuss the solution of this linear equation and derive therefrom the accuracy of the mass estimation in dependence of the parameter of a weighing experiment, being the number of masses $n$ and measurements $N$, as well as the variance of the measurement error.

As a proof of concept, we perform an experiment with a set of $n = 8$ mass elements.
Under consideration of a constant offset $\mgt_0$ as introduced in (\ref{eq:linModel}), we 
have $n+1=9$ unknowns $\{\mgt_i, i=0,\dots,n\}$, which we determine from a large set of measurements $N$.
Experimental details will be given in the following section \ref{sec:Experiment}.
Before, we exemplify the concept according to this realization.\\

Combining the $N$ measurements $y_j$ to a column vector $\mathbf{y}$ and the coefficient vectors 
$\mathbf{x}_j$ to a $N \times (n\!+\!1)$  matrix $\mathbf{X}$, 
the linear model \eqref{eq:linEqSys} can be written as a linear equation system
\begin{equation}
\label{eq:linEqSysVec}
\mathbf{y} = \mathbf{X\mgt} + \boldmath{\varepsilon}
\end{equation}
with a so-called {\em design matrix} $\mathbf{X}$.
With $n \!=\! 8$ mass elements, $N \!=\! 2^n\! =\! 256$ combinations are possible, thus resulting in a $256$ dimensional measurement vector 
$\mathbf{y} = (y_1, \ldots, y_{256})^T$.
One particular experimental realization of the linear equation system~\eqref{eq:linEqSysVec} might read as
\begin{equation}
\label{eq:xmplLinEqSys}
\begin{pmatrix}
y_1 \\
y_2 \\
y_3 \\
y_4 \\
\vdots\\
\vdots\\
\vdots\\
y_{253} \\
y_{254} \\
y_{255} \\
y_{256} \\
\end{pmatrix}
= 
\begin{pmatrix}
1 & 0 & 0 & 0 & 0 & 0 & 0 & 0 & 0\\
1 & 1 & 0 & 0 & 0 & 0 & 0 & 0 & 0\\
1 & 1 & 1 & 0 & 0 & 0 & 0 & 0 & 0\\
1 & 1 & 1 & 0 & 0 & 0 & 0 & 0 & 1\\
1 & 0 & 1 & 0 & 0 & 0 & 0 & 0 & 1\\
\vdots & \vdots & \vdots & \vdots & \vdots & \vdots & \vdots & \vdots & \vdots\\
\vdots & \vdots & \vdots & \vdots & \vdots & \vdots & \vdots & \vdots & \vdots\\
\vdots & \vdots & \vdots & \vdots & \vdots & \vdots & \vdots & \vdots & \vdots\\
1 & 1 & 1 & 0 & 1 & 1 & 1 & 1 & 1\\
1 & 1 & 1 & 0 & 1 & 1 & 1 & 1 & 0\\
1 & 1 & 1 & 1 & 1 & 1 & 1 & 1 & 0\\
1 & 1 & 1 & 1 & 1 & 1 & 1 & 1 & 1\\
\end{pmatrix}
\begin{pmatrix}
\mgt_0 \\
\mgt_1 \\
\mgt_2 \\
\mgt_3 \\
\mgt_4 \\
\mgt_5 \\
\mgt_6 \\
\mgt_7 \\
\mgt_8 \\
\end{pmatrix}
+
\begin{pmatrix}
\varepsilon_1 \\
\varepsilon_2 \\
\varepsilon_3 \\
\varepsilon_4 \\
\vdots\\
\vdots\\
\vdots\\
\varepsilon_{253} \\
\varepsilon_{254} \\
\varepsilon_{255} \\
\varepsilon_{256} \\
\end{pmatrix}.
\end{equation}
The rows of the equation system can be permuted without changing the result of the regression.
One may therefore choose a scheme for the order of the experiments, in which as less elements as possible are exchanged between two measurements \cite{Savage1997}.\\

\section{Experimental proof of concept}
\label{sec:Experiment}



\subsection{Set-up}
\label{ssec:setup}
\begin{figure}
\begin{center}
\includegraphics*[width = 0.75\textwidth]{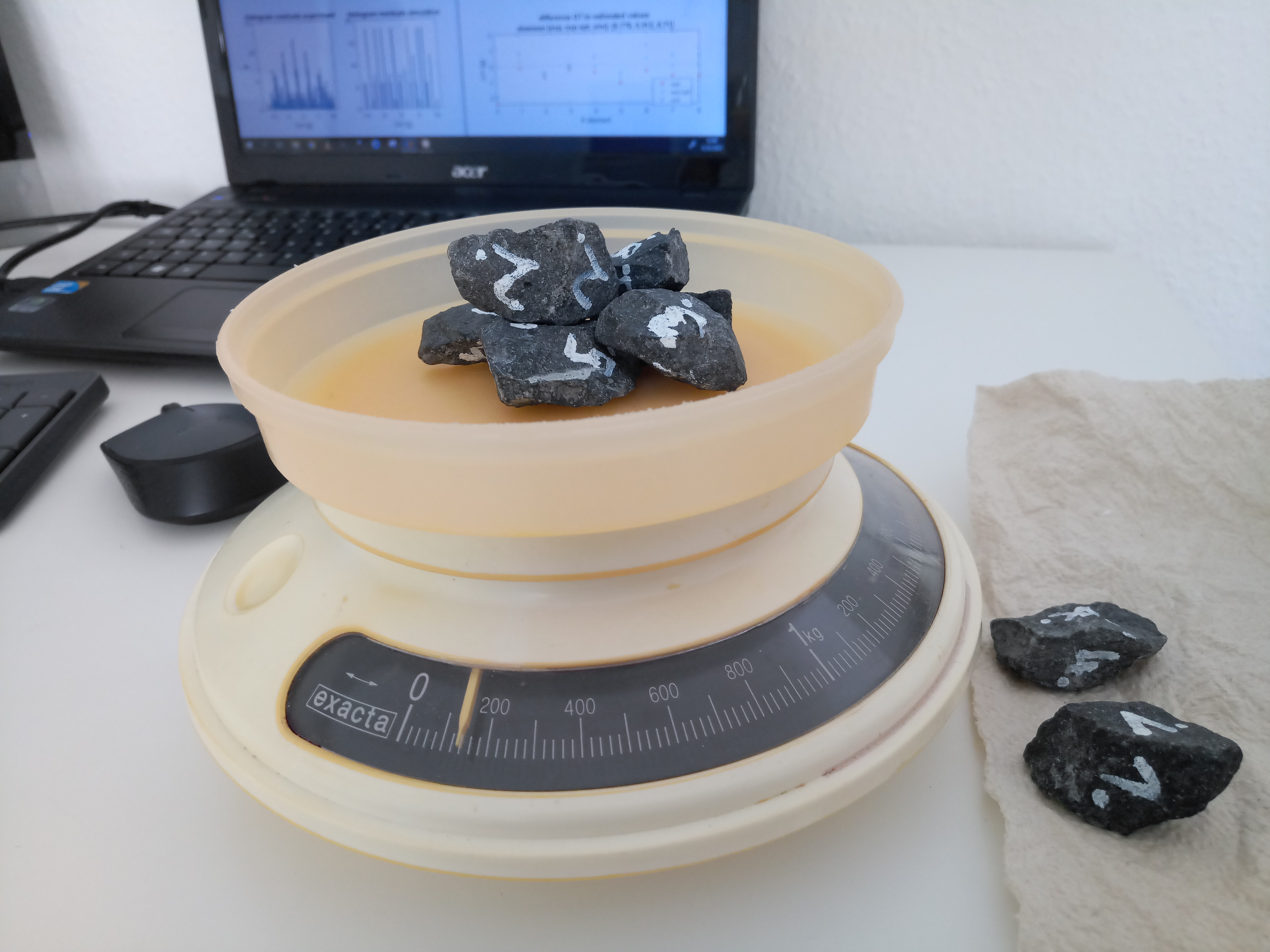}
\caption{Proof-of-concept set-up: Mechanical spring scale for kitchen purposes and numbered stones resembling the set of unknown masses.}
\label{fig:kitchenscale}
\end{center}
\end{figure}

The proof-of-concept experiments are intentionally kept simple with regard to the required equipment.
Consequently, they only incorporate a mechanical kitchen scale with a reading precision of $a = \SI{20}{\g}$, cf. figure \ref{fig:kitchenscale}, and a set of stones.
Twelve numbered stones resemble the full set of unknown masses.
In one of our measurements campaigns, we also consider a subset of eight stones.
The presented experiments were guided by a graphical user interface app which is shared in a public \textit{GitHub} repository \cite{githubRep:Reiswaage}.

Now, following the proposed concept, the task is to determine the weight of each stone as precisely as possible.
For verifying the accuracy of the result, we use a digital precision scale with a reading precision of \SI{1}{\milli\gram}.
The determined values are considered as ground truth (GT).
In section \ref{sec:Theory}, the theoretical considerations will give an estimate of how precise the method can get and what the influencing parameters are.

\subsection{Results}
\label{sec:ExpResults}
In the following sections, we will compare the estimates of each of the mass elements with the according GT values determined with the reference scale of higher precision.
However, measuring the ground truth for the offset $m_0$ (the tare) is a more complex process and cannot be done directly in a single measurement with a reference scale.
One approach is to weigh multiple reference masses and build the average difference of the known masses to the values shown by the scale in question.

In our proof of concept experiments, we will perform exactly such a measurement campaign, and additionally, we know the GT values $\mgt_i$ of the $n$ mass elements. 
We moreover assume that the rounding error (see \eqref{eq:linEqSys}) is equally distributed such that its average over $N$ measurements becomes small, i.e. 
\begin{equation*}
\frac{1}{N}\sum\limits_{j = 1}^N \left( y_j -  \mathbf{x}_j \mathbf{\mgt} \right) 
= \frac{1}{N}\sum\limits_{j = 1}^N \varepsilon_j \approx 0.
\end{equation*}
And hence, 
we can separate $m_0$ easily as it equally contributes to every measurement (see \eqref{eq:linModel} and \eqref{eq:sumWeight}), yielding
\begin{equation}
\label{eq:m0_GT}
m_0=\frac{1}{N}\sum\limits_{j = 1}^N \left(y_j - \sum\limits_{i = 1}^n x_{i,j} m_i\right).
\end{equation}
The GT values for $m_0$ presented in the following are determined in this way and marked with an asterisk in the respective tables below.

\begin{table}[h]
    \centering
    \begin{tabular}{|r|c|c|c|c|c|}
    \hline
    $i$ & $\mgt_i$ & $\miw_i$ & $\mlr_i$ & $\miw_i - \mgt_i$ & $\mlr_i - \mgt_i$\\
    \hline
    0 & -4.711* & 0 & -2.344 & 0 & 2.368\\
    1 & 36.421 & 40 & 36.250 & 3.579 & -0.171\\
    2 & 33.269 & 40 & 32.500 & 6.731 & -0.769\\
    3 & 19.998 & 20 & 19.688 & 0.002 & -0.311\\
    4 & 31.083 & 40 & 30.625 & 8.917 & -0.458\\
    5 & 28.949 & 20 & 27.813 & -8.949 & -1.137\\
    6 & 27.284 & 20 & 26.875 & -7.284 & -0.409\\
    7 & 16.664 & 20 & 15.938 & 3.336 & -0.727\\
    8 & 16.692 & 20 & 15.938 & 3.308 & -0.755\\
    \hline
    \end{tabular}
    \caption{$\mgt_i$:~Ground truth values, $\miw_i$:~individual weighing with reading precision $a = \SI{20}{\g}$, $\mlr_i$:~values determined with proposed method; all values in [g]. *: Estimate of the GT calculated according to \eqref{eq:m0_GT}.}
    \label{tab:expResult_N256}
\end{table}

\subsubsection{$N = 2^n$ weighing scheme}
\label{ssec:expTwoToN}
First, we consider a set of $n = 8$ stones, which gives $N = 2^n = 256$.
Their averaged mass is $\overline{\mgt} = \SI{23.37}{\g}$ with a standard deviation of $\sigma_{\mgt} = \SI{11.30}{\g}$.
Each of the $N = 256$ measurements $y_j$ is, of course, a multiple of the reading precision $a$. 
The particular realization  \eqref{eq:xmplLinEqSys} of a design matrix yields, e.g the measurement vector
\begin{equation*}
    \mathbf{y} = (0, 20, 40, 60, 40,\ldots, 140, 120, 160, 200)^T \si{\g}.
\end{equation*}
With a linear regression approach as described in detail in section \ref{sec:Theory}, we look for a solution $\mathbf{\mlr}$ that minimizes the square error \eqref{eq:SEdef1}.
The resulting estimates $\mlr_i$ are given in table \ref{tab:expResult_N256} together with the ground truth values $\mgt_i$ and the individual weighing results $\miw_i$.
To summarize the experimental outcome, we found a maximum difference of the linear regression results to the ground truth values of $\text{max}(|\mlr_i - \mgt_i|) = \SI{1.14}{\g}$ for $i = 1\ldots8$.
In comparison to the expected maximum difference for an individual weighing $\text{max}(|\miw_i - \mgt_i|) = a/2 = \SI{10}{\g}$, this means an increase of precision by around one order!

Moreover, the root mean square (RMS) error 
\begin{equation}
\label{eq:RMS}
    \rms_{\mlr} = \sqrt{\frac{1}{n}\sum_{i=1}^n(\mlr_i-\mgt_i)^2}
\end{equation}
of the estimated masses with respect to the GT values is $\rms_{\mlr} = \SI{0.66}{\g}$.

The negative offset $\mgt_0=\SI{-4.7}{g}$ of the scale, determined via formula \eqref{eq:m0_GT}, indicates that the scale weighs too less in comparison with the true weights. With the proposed linear regression method, this offset is also estimated, though with a lower accuracy than the $n$ masses $\mgt_i$ (compare section \ref{ssec:Theo_alln}).
Notice that all the estimated values $\mlr_i, i=1,\ldots,n$ are smaller than the corresponding ground truth values $\mgt_i$.
On the contrary, the negative offset is overestimated thus compensating for the underestimation of the masses in the linear model. 


With $N = 256$ measurements, achieving $\rms_{\mlr} = \SI{0.66}{\g}$ is a major improvement compared to the individual weighing procedure with $\rms_{\miw} = \SI{6.05}{\g}$.
Going to higher $n$ would consequently increase the effort rapidly as for increasing $n$ by one would mean a doubling of the amount of possible combinations.
It is, however, certainly not necessary to measure all combinations, and adapting the weighing scheme to the desired precision and required effort might be beneficial.

\subsubsection{Scheme with fixed number of masses on scale}
\label{ssec:expnCHk}
In this scheme the number $k$ of mass elements on the scale is fixed to a constant value. 
One obvious advantage is the smaller effort compared to the $N = 2^n$ approach and -- on top -- the scheme can be adapted by selecting a particular value for $k$. 
As one can easily see, it is possible to build up this scheme by exchanging only two masses at a time between two measurements.
For this scheme the needed dynamic range of the scale is much smaller than for performing all possible measurements.
This implies that the scale can be calibrated more accurately in a desired weight range in case it shows some non-linearities over the full range.

Later on in section \ref{sec:Theory}, we will see why it is wise to include the null-measurement, e.g. measuring once without any element (or an offset element only) on the scale.
With this additional measurement, we have a total number of combinations of $N = {n\choose{k}} + 1$.
In a particular experimental realization, we expand our set of stones with another $4$ entities to $n = 12$.
Their averaged mass is $\overline{\mgt} = \SI{24.94}{\g}$ with a standard deviation of $\sigma_{\mgt} = \SI{7.80}{\g}$.

Now choosing $k = 9$, we consequently end up with $N = 221$ measurements.
This is about the same amount of measurements as for the $N = 2^{n=8}$ experiment.
\begin{table}[]
    \centering
    \begin{tabular}{|r|c|c|c|c|c|}
    \hline
    $i$ & $\mgt_i$ & $\miw_i$ & $\mlr_i$ & $\miw_i - \mgt_i$ & $\mlr_i - \mgt_i$\\
    \hline
    0 & -3.344* & 0 & 0 & 0 & 3.344\\
    1 & 36.421 & 40 & 35.232 & 3.579 & -1.189\\
    2 & 33.269 & 40 & 32.566 & 6.731 & -0.703\\
    3 & 19.998 & 20 & 20.566 & 0.002 & 0.568\\
    4 & 31.083 & 40 & 30.343 & 8.917 & -0.740\\
    5 & 28.949 & 20 & 28.566 & -8.949 & -0.383\\
    6 & 27.284 & 20 & 27.232 & -7.284 & -0.052\\
    7 & 16.664 & 20 & 15.677 & 3.336 & -0.987\\
    8 & 16.692 & 20 & 15.677 & 3.308 & -1.015\\
    9 & 27.683 & 20 & 27.677 & -7.683 & -0.006\\
    {10} & 29.705 & 20 & 28.566 & -9.075 & -0.509\\
    {11} & 20.328 & 20 & 21.899 & -0.328 & 1.571\\
    {12} & 27.821 & 20 & 26.788 & -7.821 & -1.033\\
    \hline
    \end{tabular}
    \caption{$\mgt_i$:~Ground truth values, $\miw_i$:~individual weighing with reading precision $a = \SI{20}{\g}$, $\mlr_i$:~values determined with proposed method; all values in [g]. *: Estimate of the GT calculated according to \eqref{eq:m0_GT}.}
    \label{tab:expResult_12CH9}
\end{table}
Results for each mass element are presented in table \ref{tab:expResult_12CH9}.
Here, we find a RMS of $\rms_{\mlr} = \SI{0.85}{\g}$ with a maximum difference to the ground truth of \nolinebreak{$\text{max}(|\mlr_i-\mgt_i|) = \SI{1.57}{\g}$}. 
With a comparable effort (amount of measurements) as the $N = 2^n$ experiment, we find the RMS value around $\SI{30}{\percent}$ less precise than the estimates found in section \ref{ssec:expTwoToN}.
In turn, we have estimated the weights of four additional masses!
The estimated offset $\mgt_0$ is, however, identical to the zero measurement of the scale, as discussed in \ref{ssec:Theo_nChk+1}. This means that for this weighing scheme the offset accuracy is not increased.

\subsection{Experimental résumé}
The initial question of whether the reading precision of a scale can be under-run with the proposed combined measurements method can definitely be answered with "yes".
We have seen that the two weighing schemes reduce the precision range drastically by more than one order.
However, further questions related to the obtainable precision or crucial influencing factors arise immediately.
To give answers more quantitatively, we will analytically examine the proposed method in the following section before presenting results of more general experimental simulations and statistics.

\newpage
\section{Analytical considerations}
\label{sec:Theory}
\subsection{Basic mathematical concept}
The concept of linear regression emerges already more than 200 years ago with Legendre and Gau{\ss} (see e.g. \cite{wikipedia:Regression} and references therein). 
A modern representation of multiple linear regression can, e.g.,~be found in \cite{Freedman:2009}. We recommend also the thorough and comprehensive representation of linear models for regression and classification in the context of machine learning by Bishop \cite{bishop:2006:PRML}.
Here we very briefly recapitulate the basic mathematical concept in order to be able to investigate analytically the dependence of the regression accuracy on the number of masses $n$ and measurements $N$ and to understand the limits of the approach.

The so-called multiple regression assumes a linear measurement model 
\begin{equation}
\label{eq:linEqSysVec_rep}
\mathbf{y} = \mathbf{X\mgt} + \boldmath{\varepsilon}
\end{equation}
which has already been introduced for the weighing process in section \ref{sec:concept}. 
Thereby $\mathbf{X}$ is the $N \times (n\!+\!1)$ so-called {\em design matrix} of $N$ measurements and $\mathbf{y}$ a column vector that combines the $N$ measurement results $y_j$\,.

Linear regression estimates the $(n\!+\!1)$ coefficients $\mathbf{m}=(m_0,\ldots,m_n)$ by minimizing the square error \eqref{eq:SEdef1} which can be written in matrix notation as
\begin{align}
\label{eq:SEdef2}
E = (\mathbf{y}-\mathbf{Xm})^T(\mathbf{y}-\mathbf{Xm}).
\end{align}
The standard way to find the minimum of \eqref{eq:SEdef2} is to take the derivative with respect to $m$ and setting it to zero, i.e.
\begin{equation}
\begin{aligned}
\label{eq:zero}
\mathbf{X}^T(\mathbf{y}-\mathbf{X}\mathbf{\hat{m}}) &= 0 \\
  \Leftrightarrow \quad \mathbf{X}^T\mathbf{X} \mathbf{\hat{m}} &= \mathbf{X}^T\mathbf{y}.
\end{aligned}
\end{equation}
The solution of this normal equation yields the estimated coefficients $\mathbf{\hat{m}}$ as
\begin{equation} 
\label{eq:wsol}
\mathbf{\hat{m}}  = \mathbf{A}^+ \mathbf{X}^T\mathbf{y}
\end{equation}
with $\mathbf{A}^+$ being the inverse of the symmetric $(n+1)\times (n+1)$ matrix 
\begin{equation}
\label{eq:Adef}
\mathbf{A} = \mathbf{X}^T\mathbf{X}.
\end{equation}
In case $\mathbf{A}$ is singular, i.e.~having not the full rank $n\!+\!1$, one can still compute the so-called {\em Moore-Penrose inverse} (also called {\em pseudo inverse}) of $\mathbf{A}$ (see \cite{Ben-Israel:2003,wikipedia:Moore-Penrose}).
The solution provided by \eqref{eq:wsol} may, however, then not be the unique solution of (\ref{eq:zero}).

\subsubsection{Accuracy of the regression}
\label{se_ep}
The fundamental question we treat in this and in the next subsection is how accurate linear regression can estimate the coefficients of a linear model.
The {\em Gau{\ss}-Markov} theorem states that minimizing the square error \eqref{eq:SEdef2} yields the best linear unbiased estimate, if the errors in the linear measurement model \eqref{eq:linEqSysVec_rep} are uncorrelated, have equal variances, and an expectation value of zero, i.e.
\begin{align}
\label{eq:GM-condition1}
\langle \boldsymbol{\varepsilon} \rangle &= \mathbb{0}\\
\label{eq:GM-condition2}
\mathrm{cov}(\boldsymbol{\varepsilon}) &= \sigma^2\,\mathbb{1}
\end{align}
with $\mathbb{1}$ being the $(n\!+\!1)\!\times\!(n\!+\!1)$ identity matrix and $\mathbb{0}$ the $(n\!+\!1)$ dimensional vector of zeros.
The best linear unbiased estimate means thereby the estimate $\hat{\mathbf{m}}$ of the model parameter with the smallest sampling variance around the true parameter ${\mathbf{m}}$. 

In appendix \ref{sec:ycor}, we identify the condition under which the {\em Gau{\ss}-Markov} theorem is fulfilled for the case that the measurement error originates from a rounding error.
That is, one can assume a mean value free rounding error with constant variance, as required in equations \eqref{eq:GM-condition1} and \eqref{eq:GM-condition2}, 
if the standard deviation $\sigma_m$ of the ground-truth masses $m_i$ is larger than half of the scale resolution $a$, e.g.
\begin{equation} 
\label{eq:GMcondition} 
\sigma_m > \frac{a}{2}.
\end{equation}

In case the measure error variances are not equal, one can still obtain an optimal estimate by performing a weighted least square fit. 
Therefore, one introduces a diagonal $N \times N$ matrix in the error function \eqref{eq:SEdef2} in which each diagonal element corresponds to the inverse of the respective measurement variance \cite{Wiora2016,Ho2020,Montgomery2012}. 
An even more general approach that also takes error correlations into account is the {\em Gau{\ss}-Markov} method.
This method provides a minimum variance, linearly unbiased estimator rather than a minimum least square estimator \cite{Kruskal_1968,Bich_1990}.

To obtain an expression for the error in the estimated coefficients, one inserts the model
\eqref{eq:linEqSysVec_rep} into the solution (\ref{eq:wsol}) yielding
\begin{equation} 
\label{eq:error_m}
\hat{\mathbf{m}} = \mathbf{m} +
\mathbf{A}^+ \mathbf{X}^T\, \boldsymbol{\varepsilon}\,
\end{equation}
This expression shows that the estimate $\mathbf{\hat m}$ is bias free, i.e.~$\langle \mathbf{\hat m} \rangle = \mathbf{m}$, if the error $\boldsymbol{\varepsilon}$ is mean value free, as required by the {\em Gau{\ss}-Markov} theorem.
Therefore, it is wise to incorporate a constant term in the linear model as described in section \ref{sec:concept} to account for eventual systematic offsets in the measurement errors, which otherwise would bias the linear regression. 

The covariance of the estimate $\mathbf{\hat m}$ is 
\begin{equation} 
\label{eq:sem1}
\mathrm{cov}({\mathbf{\hat m}}) =
\mathbf{A}^+ \mathbf{X}^T \mathrm{cov}(\boldsymbol{\varepsilon}) \mathbf{X}\mathbf{A}^+.
\end{equation}
If the second condition \eqref{eq:GM-condition2} of the {\em Gau{\ss}-Markov} theorem is fulfilled,
the noise covariance matrix commutes with the other matrices and \eqref{eq:sem1} takes thus the simple form
\begin{equation} 
\label{eq:sem2}
\mathrm{cov}({\mathbf{\hat m}}) = 
\sigma^2 \mathbf{A}^+ \mathbf{A}  \mathbf{A}^+ = \sigma^2 \mathbf{A}^+.
\end{equation}
Note that the later equation uses explicitly one of the definitions of a pseudo-inverse \cite{Ben-Israel:2003, wikipedia:Moore-Penrose}.


\subsection{Analytical treatment of the regression error}
\label{sec:error_calc2n}
Formula \eqref{eq:sem2} expresses the relation (\ref{eq:sem1}) between the expected regression error of the estimated coefficients (in our cases the masses) and the measurement errors, in which the the matrix $A$ plays the role of an inverse scale factor.
This matrix, which counts the correlated occurrences of the $n$ different masses, can in practice for each experiment be easily calculated from the design matrix of the experiment using definition (\ref{eq:Adef}).

While this formula is known in the state of the art (see e,g,~\cite{Montgomery2012,Freedman:2009}) we here go further and gain a deeper insight into the dependencies of the accuracy of the estimated masses on the number of measurements and contributing masses. 
For the analytic treatment we restrict ourselves to the case that all masses $m_i, i\!=\!1,\ldots,n$ occur with equal probability $p$ in the $N$ measurement combinations, i.e.\ they occur $pN$ times, while the offset $m_0$ occurs always. 
Assuming furthermore that the joint probability for two different masses to occur together is equal to $q$ for all pairs of masses, the $(n+1)\times(n+1)$ matrix $A$ takes the special form
\begin{equation}
\label{eq:AN}
\mathbf{A}=N
\begin{pmatrix}
1 & p & \cdots &\cdots & p\\
p & p & q & \cdots & q\\
\vdots & q & \ddots & \ddots & \vdots\\
\vdots & \vdots & \ddots & \ddots & q\\
p & q & \cdots & q & p
\end{pmatrix}.
\end{equation}

The inverse $A^+$ of a matrix $A$ exists if its rows are linearly independent. 
In this case the inverse, being the solution of the linear equation system
\begin{equation}
\label{eq:AAinv}
AA^+=\mathbb{1}\,,
\end{equation}
can be calculated with the well-known {\em Gau\ss-Jordan} algorithm, in which the matrix $A$ is brought into a diagonal form by subsequent linear combinations of the rows of the linear equation system. 
Applying the same transformations to the identity matrix yields the inverse matrix.

In order to obtain the general form of the inverse  of $A$, we make the following ansatz
\begin{equation}
\label{eq:AN_inv}
\mathbf{A^+}=\frac{1}{N}
\begin{pmatrix}
\alpha & \beta & \cdots &\cdots & \beta\\
\beta & \gamma & \delta & \cdots & \delta\\
\vdots & \delta & \ddots & \ddots & \vdots\\
\vdots & \vdots & \ddots & \ddots & \delta\\
\beta & \delta & \cdots & \delta & \gamma\\
\end{pmatrix}
\end{equation}
with four unknowns $\alpha, \beta, \gamma, \delta$. With this ansatz the linear equation system \eqref{eq:AAinv} is reduced to a set of only $5$ linear equations, which are
\begin{equation}
\label{eq:AAinv5}
    \mathbb{1}_{i,j} = \left\{
\begin{aligned}
    \alpha + n p \,\beta                &= 1 \quad \mathrm{for} \quad i=j=1 \\
    p\alpha + [p+(n-1) q]\beta           &= 0 \quad \mathrm{for} \quad i>1, j=1\\
    \beta + p\gamma +(n-1)p\delta       &= 0 \quad \mathrm{for} \quad i=1, j>1 \\
    p\beta + p\gamma + (n-1)q\delta      &= 1 \quad \mathrm{for} \quad i=j>1 \\
    p\beta + q\gamma +[p+(n-2)q]\delta  &= 0 \quad \mathrm{for} \quad i\neq j;\, i,j>1
\end{aligned}
\right.
\end{equation}
or, written as linear equation system with the unknown parameters as vector $(\alpha, \beta, \gamma, \delta)^T$:
\begin{equation}
\label{eq:AAinv5M}
\begin{pmatrix}
1 & np & 0 & 0\\
p & p\!+\!(n\!-\!1)& 0 & 0\\
0 & 1 & p & (n\!-\!1) p\\
0 & p & p & (n\!-\!1) q\\
0 & p & q & p\!+\!(n\!-\!2) q\\
\end{pmatrix}
\begin{pmatrix}
\alpha\\
\beta\\
\gamma\\
\delta
\end{pmatrix}
=
\begin{pmatrix}
1\\
0\\
0\\
1\\
0
\end{pmatrix}
.
\end{equation}
Equation \eqref{eq:AAinv5M} can again be solved with the {\em Gau\ss-Jordan} algorithm with the result
\begin{align}
\label{eq:Ainv5M}
\begin{split}
\alpha &= 1+\frac{np^2}{Q}\\
\beta &= -\frac{p}{Q}\\
\gamma &= \frac{1}{Q}\left(1-(n\!-\!1)\frac{p^2-q}{p-q}\right)\\
\delta &= \frac{1}{Q}\frac{p^2-q}{p-q}\\
\end{split}\\
\mathrm{with}\qquad &Q = p - np^2+(n\!-\!1)q.
\label{eq:Q}
\end{align}

The normalization factor $Q$ which occurs in all $4$ parameters in a denominator has been introduced for convenience. 
The step-by-step derivation of this solution using the {\em Gau\ss-Jordan} algorithm is presented in appendix A.

One recognizes immediately that the solution
(\ref{eq:Ainv5M}) becomes singular for $Q=0$. We have thus derived a simple criteria which tells us when the matrix (\ref{eq:AN}) is singular. 
In this case, the pseudo-inverse may be computed, but which does not provide a unique solution, unless the matrix $A$ has full column rank  \cite{Ben-Israel:2003, wikipedia:Moore-Penrose}.

With the five parameters provided by \eqref{eq:Ainv5M} and \eqref{eq:Q}, the inverse $A^+$ \eqref{eq:AN_inv} of the special matrix $A$ \eqref{eq:AN} is fully determined. 
Thereby, according to formula \eqref{eq:sem2}, the two diagonal values $\alpha$ and $\gamma$ are proportionality factors between the measurement noise $\sigma$ and the expected errors in the estimated coefficients $\hat{m}_i$, which scale also inverse with the total number $N$ of experiments i.e.
\begin{align}
\begin{split}
\label{eq:delta_m} 
\text{var}(\mlr_0) &=\alpha \,\frac{\sigma^2}{N}\\
\text{var}(\mlr_i) &= \gamma \,\frac{\sigma^2}{N}, 
\quad \text{for} \quad i=1,\ldots,n.
\end{split}
\end{align}
This result will in the following be applied to different cases of experiment designs.

\subsubsection{Example 1: The case $N=2^n$}
\label{ssec:Theo_alln}
As discussed above, in the case that all $N=2^n$ combinations of the $n$ masses are weighed, each weight occurs $N/2$ times and each combination of two weights $N/4$ times.
This implies for the probabilities in the matrix (\ref{eq:AN}) $p=1/2$ and $q=1/4$.

Inserting these values in \eqref{eq:Q} one obtains $Q=1/4=p^2=q$ and thus from \eqref{eq:Ainv5M}
\begin{align}
\begin{split}
\alpha &= 1+n\\
\beta &= -2\\
\gamma &= 4\\
\delta &= 0\, .
\end{split}
\end{align}
Considering these values for 
$\alpha$ and $\gamma$ of $A^+$ in \eqref{eq:delta_m} yields
\begin{eqnarray}
\label{eq:thMdl_2TOn_m0}
\text{var}(\mlr_0) &=& \frac{(1+n)\,\sigma^2}{N}\\
\label{eq:thMdl_2TOn_mi}
\text{var}(\mlr_i) &=& \frac{4\,\sigma^2}{N}\,, \qquad\mathrm{for}\quad i = 1 \ldots n.
\end{eqnarray}
One recognizes that the variances of the estimated masses decay inversely with the number of experiments $N$. 
The dependencies are visualized in figure \ref{fig:simDelmVSn}, below in section \ref{sec:Simulations}.
%
%
%
\subsubsection{Example 2: The case $N = {n\choose k}$}
\label{AppSec:nChk_expl}
In case that one weighs all possible combinations of $k$ mass elements chosen from $n$, one performs in total a number of 
\begin{equation}
    N = {{n}\choose{k}} = \frac{n!}{k! \, (n\!-\!k)!}
\end{equation}
weight measurements.
In order to determine the occurrence probabilities $p$ and $q$ in matrix $A$ (\ref{eq:AN}), one needs to determine the number of combinations in which a certain mass element, respectively a combination of two mass elements, occurs.
Selecting one mass element out of $n$, leaves $(n\!-\!1)$ elements to choose the other $(k\!-\!1)$. 
And selecting two elements out of $n$ leaves $(n\!-\!2)$ elements to choose the remaining $(k\!-\!2)$. 
One obtains thus
\begin{align}
p &= \frac{1}{N}{{n\!-\!1}\choose{k\!-\!1}}={{n}\choose{k}}^{-1}{{n\!-\!1}\choose{k\!-\!1}} = \frac{k}{n} \\
q &= \frac{1}{N}{{n\!-\!2}\choose{k\!-\!2}}={{n}\choose{k}}^{-1}{{n\!-\!2}\choose{k\!-\!2}} = \frac{k(k\!-\!1)}{n(n\!-\!1)}.
\end{align}
Inserting $p$ and $q$ in (\ref{eq:Q}), one finds $Q\!=\!0$ which means that the matrix $A$ (\ref{eq:AN}) is singular and thus there is no unique solution.
One quickly realizes that the $n \choose k$ case is under-determined for $k = 1$ or $k = n$ as, with the scale offset $m_0$, there are  $n\!+\!1$ parameters in the linear equation system rather than $n$.
However, even for $1\!<\!k\!<\!n$, there is no unique solution $\mathbf{\mlr}$ minimizing the error \eqref{eq:SEdef2}. One sees easily that if $(\mlr_0,\mlr_1,\ldots,\mlr_n)$ 
minimizes the error \eqref{eq:SEdef1}, also $(\mlr_0\!+\!\delta_0,m_1\!-\!\delta_0/k,\ldots,m_n\!-\!\delta_0/k)$ does for any offset
$\delta_0$, because each measurement involves the same number $k$ of mass elements.

\subsubsection{Example 2a: The case $N = {n\choose k} + 1$}
\label{ssec:Theo_nChk+1}
The singularity in $A$ \eqref{eq:AN}
can be overcome by considering an additional null-measurement $y_0$ without any elements on the scale such that one ends up with $N = {{n}\choose{k}}+1$ measurements in total.
The additional term $(y_0\!-\!m_0)^2$ in the error \eqref{eq:SEdef1} is obviously minimal for $\mlr_0\!=\!y_0$,
and in this way, the above discussed ambiguity in the solution $\mathbf{\mlr}$ is resolved.

With similar arguments as in \ref{ssec:Theo_alln}, one obtains
\begin{align}
\begin{split}
p &= \frac{1}{N}{{n\!-\!1}\choose{k\!-\!1}}= \frac{k (N\!-\!1)}{nN}, \\
q &= \frac{1}{N}{{n\!-\!2}\choose{k\!-\!2}}= \frac{k(k\!-\!1)(N\!-\!1)}{n(n\!-\!1)N}, \\
Q &= \frac{k^2(N\!-\!1)}{nN^2}.
\end{split}
\end{align}
Since $Q\!\neq\!0$ for $N > 1$, analytical expressions for the $A^+$ elements can be found, which are
\begin{align}
\begin{split}
\alpha & = N, \\
\beta &= -\frac{N}{k},\\
\gamma &= \frac{N}{k(N\!-\!1)}\left(\frac{N}{k} + \frac{(n\!-\!1)^2}{(n\!-\!k)}\right),\\
\delta &= \frac{N}{k(N\!-\!1)}\left(\frac{N}{k} - \frac{(n\!-\!1)}{(n\!-\!k)}\right).
\end{split}
\end{align}
Inserting these expressions in \eqref{eq:delta_m} one  obtains as error variances
\begin{eqnarray}
\label{eq:thMdl_nCHk_m0}
\text{var}(\mlr_0) &=& \sigma^2\qquad \qquad  \text{and}\\
\label{eq:thMdl_nCHk_mi}
\text{var}(\mlr_i) &=& \frac{\sigma^2}{k(N\!-\!1)}\left(\frac{N}{k} + \frac{(n\!-\!1)^2}{(n\!-\!k)}\right)
\end{eqnarray}
for $i = 1 \ldots n$. Remarkable is the fact that the error in the estimated offset $\mlr_0$ equals the measurement error $\sigma$ independently of the number of measurements, 
because the estimated offset equals the null measurement.
In contrast, the error variance of the estimated masses 
$\mlr_i$  shows a rather complex dependency on $k$, as illustrated in the theoretical curves plotted in figure~\ref{fig:sim_nCHk} below in section \ref{sec:Simulations}.

\section{Statistical simulations}
\label{sec:Simulations}
In order to verify the results of the theoretical part, we need to make statistical statements.
With regard to the number of measurements and comprehensive simulation campaigns to be performed, experimental verification would be an enormous effort.
But instead, we build up a system of experimental simulations to gain the necessary statistical significance.


\subsection{General framework}
The basis for the statistical simulations is a randomized set of mass elements as a ground truth input where the number, average, and standard deviation of normally distributed masses are the only parameters.
For instance, we generate a ground truth set of $n = 8$ masses with the average and standard deviation values given by our experiment. 
Additionally, the scale offset $m_0$ is randomized uniformly within the reading precision interval $]\!-\!a/2, a/2]$. This implies that we assume for the zero measurement the same accuracy as for any other measurement. A more precise taring of the scale could also be simulated and accounted for in the theory within the framework of weighted least square fit, 
briefly discussed in section \ref{se_ep}, where the zero measurement would get a higher weight than the $N-1$ other measurements.

Having generated such a random set of GT, we simulate $N$ weighing processes.
That is, for each $j$ of all $N$ measurements, building the ground truth total weight from this input set $\mathbf{\mgt}$ and a particular binary mass-selection vector $\mathbf{x}_j$ before rounding it through the predefined reading precision $a$, i.e.
\begin{eqnarray}
y_j = a\left[\mathbf{x}_j \cdot \mathbf{\mgt}/a\right],
\label{eq:rounding}
\end{eqnarray}
where $[\cdot]$ denotes the nearest integer.
Such sets of identical parameters are drawn $P = 1000$ times in order to apply our linear regression approach with statistically varying prerequisites, receiving $\mathbf{\mlr}_l,\,l\!=\!1,\!\ldots,\! P$.

For each experiment, we compute the squared RMS error of the estimated masses, $\rms_{\mlr}^2$, according to \eqref{eq:RMS}, and average over all $P$ experiments, yielding
\begin{align}
\label{eq:statRMSmasses}
  \srms^2_{\mlr} &= \frac{1}{P}  \sum_{l=1}^P \frac{1}{n} \sum_{i=1}^n
  (\mlr_{i,l} - \mgt_{i,l})^2\\
\label{eq:statRMSoffset}  
  \srms^2_{\mlr_0} &= \frac{1}{P}  \sum_{l=1}^P   (\mlr_{0,l} - \mgt_{0,l})^2,
\end{align}
to receive numerical values for the variance of the estimated masses and offset, which we compare with the theoretical values $\text{var}(\mlr_i)$ calculated from equations (\ref{eq:delta_m}).
The statistical evidence we gain will allow us to judge about the correctness of our theoretical assumptions.

\subsection{Case $N = 2^n$}
In a first simulation campaign, we have a look at the feasible precision depending on the number of mass elements where all possible $N = 2^n$ measurements were performed.
Therein, we step-wise increase the number of considered mass elements $n$ from $1$ to $12$.
\begin{figure}[t]
\begin{center}
\includegraphics*[width = 1\textwidth]{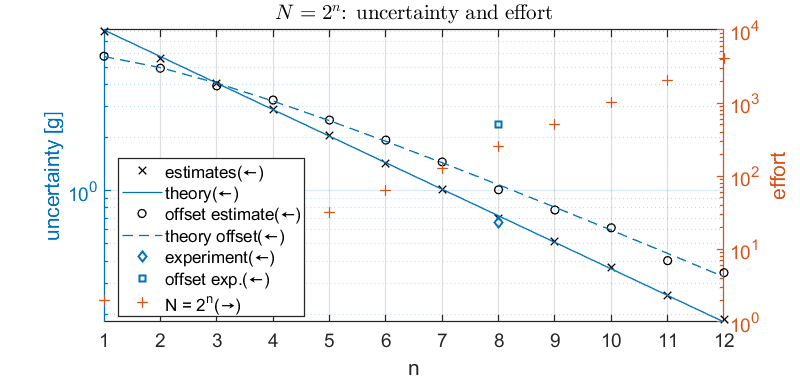}
\caption{Left axis (log scale): Uncertainties $\srms_{\mlr}$ against the number of total mass elements $n$. 
Black crosses represent the mean errors of the simulated mass estimates, black circles give the offset errors. 
Theory curves for mass (straight line) and offset mass (dashed) errors in blue. 
Blue shapes show experimental data from sec. \ref{sec:ExpResults}.
Right axis (log scale): Number of performed measurements $N$ for $n$ mass elements representing the effort.}
\label{fig:simDelmVSn}
\end{center}
\end{figure}

In figure \ref{fig:simDelmVSn}, simulated data points are presented for $\srms_{\mlr_0}$ and $\srms_{\mlr}$ from equations \eqref{eq:statRMSmasses} and \eqref{eq:statRMSoffset} against the number of mass elements $n$.
These data, which excellently match the dashed and solid lines of the theoretical predictions from equations \eqref{eq:thMdl_2TOn_m0} and \eqref{eq:thMdl_2TOn_mi}, refer to the log scaled left-hand ordinate and are labelled as uncertainty.
In addition, the effort as amount of performed measurements $N$ is presented along the right-hand ordinate.

The presented data nicely indicate an exponential decrease of the errors $\srms_{\mlr_0}$ and $\srms_{\mlr}$, going along with an exponential growth of the effort $N$.
Thus, we achieve an error of around $\srms_{\mlr} = \SI{0.18}{\g}$ for $n = 12$ with $2^{12}$ measurements performed.
Remarkable is that compared to $\srms_{\mlr}$ the offset error $\srms_{\mlr_0}$ is smaller for very few mass elements, but decays slower with $n$, such that it becomes relatively larger for more than $n = 3$ elements considered.

In agreement with the results of our experiment with $n = 8$, $\rms_{\mlr} = \SI{0.66}{\g}$ (cf. blue diamond shape in Fig. \ref{fig:simDelmVSn}), we find the average error of $\srms_{\mlr}$ approximately at $\SI{0.72}{\g}$.
However, the offset value of the experiment, indicated as blue square in the figure, seems too high to conform to the simulation and theory data.
This deviation can be understood as a sampling fluctuation in the regression error depending on the (random) choice of the $n$ masses in the specific experiments.
In the simulation, this sampling fluctuation has been eliminated by averaging over many experiments with randomly sampled masses.

\subsection{Case $N = {n\choose k}+1$}
In another simulation campaign, we consider the case of exactly $k$ of $n$ elements on the scale.
The null measurement with empty scale pan completes a set of measurements.
Adding the null measurement to the set prevents singularity, as discussed in section \ref{AppSec:nChk_expl}.

\begin{figure}[ht]
\begin{center}
\includegraphics*[width = 1\textwidth]{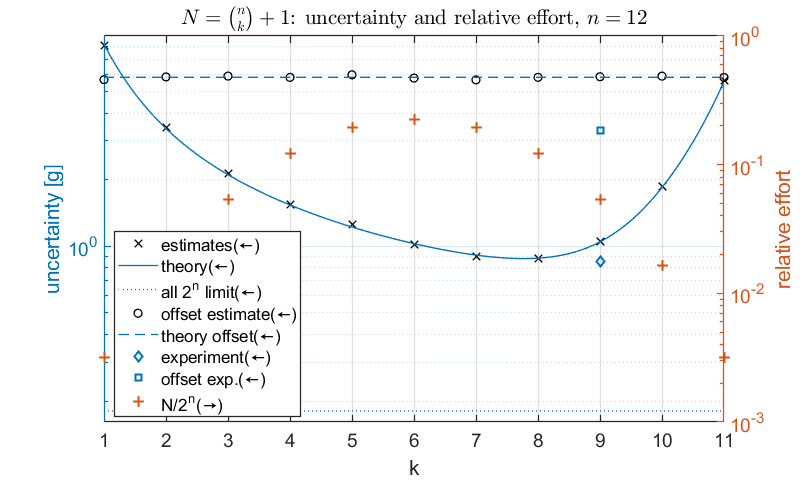}
\caption{Left axis (log scale): Uncertainties $\srms_{\mlr}$ against the number $k$ of mass elements on scale for $n = 12$.
Black crosses represent the mean errors of the simulated mass estimates, black circles give the offset errors.
Theory curves for mass (straight line) and offset mass (dashed) errors in blue. 
Dotted blue line indicates mass element error for all $2^n$ measurements. 
Blue shapes show experimental data from sec. \ref{sec:ExpResults}. 
Right axis (log scale): Relative effort in terms of number of performed measurements $N$ over the number of all possible measurements $2^n$.}
\label{fig:sim_nCHk}
\end{center}
\end{figure}
Figure \ref{fig:sim_nCHk} represents the case $n = 12$ with errors $\srms_{\mlr_0}$ and $\srms_{\mlr}$ (cf.\,\eqref{eq:statRMSmasses} and \eqref{eq:statRMSoffset}), labelled as uncertainty and plotted against $k$.
The data points from the simulations again match excellently the theoretical curves calculated from equations \eqref{eq:thMdl_nCHk_m0} and \eqref{eq:thMdl_nCHk_mi}.
As an orientation, the blue dotted line in figure \ref{fig:sim_nCHk} indicates the uncertainty for the $2^n$ case as the lower limit.

Besides a constant error of the offset element at $\srms_{\mlr_0} = \SI{5.8}{\g}$, we find a minimal error at $k = 8$ with $\srms_{\mlr} = \SI{0.9}{\g}$. 
This error is only slightly smaller than $\SI{1}{\g}$ for $k = 9$.
However, the $221$ performed measurements for $k=9$ are significantly less compared to $496$ that have to be taken for $k = 8$.

Simulation and theory yield slightly larger values for the regression error $\srms_{\mlr}$ compared to the experimental result we found in section \ref{ssec:expnCHk} with $\rms_{\mlr} = \SI{0.85}{g}$ (depicted as blue diamond shape in the figure). 
Again, this emerges from statistical fluctuations as we only compare one single experiment against a set of $P = 1000$ simulated experiments for each $n$.
We also find such a fluctuation among the experimental offset value (blue square), though having here a smaller error than expected by simulation and theory.


The asymmetry of the mass elements' uncertainty curve is remarkable as~-- due to the underlying similarity of the 
binomial coefficients to \textit{Pascal}'s triangle -- one would intuitively assume a symmetric distribution, as well.
We find such a symmetric distribution, e.g., for the effort that is presented additionally on the right-hand-side axis.
Here, the effort is the ratio of 
$N\!=\! {n\choose{k}}\!+\!1$ over $2^n$.

An explanation for the observed asymmetric uncertainty is that an error $\delta\!=\!\mlr_0\!-\mgt_0$ in the estimated offset is in the regression compensated by a bias in the other estimated masses $\mlr_i$. As discussed in section \ref{AppSec:nChk_expl} this bias is proportional to $\delta/k$ which is illustratively clear if one considers that in all measurements except the null measurement the same number $k$ of mass elements are involved.
As a result, this bias is relatively small for a larger number $k$ compared to a small number.

\section{Summary and conclusion}
In this paper, we initially discussed the basic concept of the combined measurements method by introducing the nomenclature, fundamental equations, and the approach of error minimization via linear regression estimation.

We further performed an experimental proof of concept, where we estimated the weight of stones as exemplary mass elements by two weighing schemes of different levels of experimental effort and estimation accuracy.
The accuracy of the estimates under-run the reading precision of the scale in use by one order of magnitude related to the ground-truth values.

In a next step, we summarized the mathematical basics of linear regression by least square fitting and derived a generic expressions for the covariance matrix of the estimated parameters.
This expression reveals the dependence of the regression accuracy on the set size of mass elements and number of experiments.
We subsequently gave particular analytic expressions for the regression errors that motivated experimental simulations to establish founded statistical statements.

Finally, we picked up the two weighing schemes from a comparing point of view between experiment, theory, and comprehensive simulations and found excellent agreement of the proposed concepts from all three perspectives.\\

The initial question of the feasibility to weigh individual rice corns with a truck scale remains to be answered.
It turned out that the \textit{Gau\ss-Markov} theorem provides conditions that need to be considered for answering this question.
We found, based on the considerations in the appendix \ref{sec:ycor}, that the conditions of the theorem are fulfilled if the variance of the weights is larger than a quarter of the square of the scale precision.
If this condition is not fulfilled, correlations in the measurement errors occur.
These may bias the regression result and lead to deviations not consistent with the theoretical statements for the achievable regression accuracy derived in this work. 

The above condition is obviously violated if applied to the weighing of a single rice corn. 
Nonetheless, we see a certain chance to shift the limits of this condition by introducing combined weighing schemes, as such combinations can show larger variances.
We can, however, hardly judge currently if the rice-truck scale experiment could be successful. 
Therefore future work is clearly indicated to investigate the limits of the \textit{Gau\ss-Markov} theorem for combined weighing schemes.\\

With a very simple set-up, we gained impressive results to document that weighing uncertainties limited by the scale's reading precision can be drastically reduced by magnitudes of order.
The only price to pay is an increased effort to stringently process a weighing scheme of choice, and its the choice of the experimenter of how complex the scheme is to gain a certain accuracy of estimates.
The novel analytic expressions for the expected accuracy of the combined weighing scheme allows to design optimal measurement schemes for the application under consideration, and to estimate the required effort therefore.\\

To guide the conductance of such a potentially highly complex series of combined measurements as presented, a graphical user interface was developed and shared in a public \textit{GitHub} repository.
The app is open source and as such provided as installation file within \textit{MATLAB}, and in a future release as \textit{Python} based stand-alone install.
Within this desktop app, the user can select among diverse weighing schemes.
Besides the two presented binary schemes for a total number of measurements $N \!=\! 2^n$ or $N \!=\!  {n \choose k}\!+\!1 $, also a ternary scheme is available, where the design matrix offers three different states for each mass element: $\{-1, 0, +1\}$.
This ternary scheme simulates a classical beam balance with two weighing pans on opposite sides of the beam, indicated by the coefficients $\{+1,-1\}$. This type of balance was considered in early publications on mass metrology mentioned in the introduction.

The intention of the repository is to share data sets of inspiring and creative realizations -- why not overstepping the field of weight determination?
Besides improving the app in community work, the repository should also give room to discuss -- and maybe marvel -- about experiments and their results.
An active contribution to this platform by the reader or interested person would be gratefully appreciated by the authors.

\appendix
\section{Inverse calculation by  {\em Gau\ss-Jordan} algorithm}
\label{sec:Gauss-Jordan}
The well-known {\em Gau\ss-Jordan} elimination algorithm subsequently transforms the so-called augmented coefficient matrix of a linear equation system by a series of row operations until the coefficient matrix becomes the unity matrix.  
This series of row operations for the linear equation system (\ref{eq:AAinv5M}) is shown step-by-step in the following table. 
The right column of the transformed matrix is then the solution (\ref{eq:Ainv5M}).  
The fact that the last element of this column is zero shows that the linear equation system is solvable.

\newcommand{\thvrule}{\vrule width 0.6pt \hspace*{0.2pt}}
\begin{equation*}
\label{tab:GM}
\begin{array}{@{}ccccc|cc@{}}
{\bigl \lceil} & 1 & np & 0 & 0 & 1 & {\bigl \rceil}\\
\thvrule & p & p\!+\!(n\!-\!1)q& 0 & 0 & 0 & \hspace{0.5pt} \thvrule\\
\thvrule & 0 & 1 & p & (n\!-\!1) p & 0 & \hspace{0.5pt} \thvrule\\
\thvrule & 0 & p & p & (n\!-\!1) q & 1 & \hspace{0.5pt} \thvrule\\
{\bigl \lfloor} & 0 & p & q & p\!+\!(n\!-\!2) q & 0 & {\bigl \rfloor}\\
\\[-3pt]
\hline\\[-3pt]
{\bigl \lceil} & 1 & np & 0 & 0 & 1 & {\bigl \rceil}\\
\thvrule & 0 & p\!+\!(n\!-\!1)q -np^2& 0 & 0 & -p & \hspace{0.5pt}\thvrule\\
\thvrule & 0 & 1 & p & (n\!-\!1) p & 0 & \hspace{0.5pt} \thvrule\\
{\bigl |} & 0 & p & p & (n\!-\!1) q & 1 & \hspace{0.5pt} \thvrule\\
{\bigl \lfloor} & 0 & p & q & p\!+\!(n\!-\!2) q & 0 & {\bigl \rfloor}\\
\\[-3pt]
\hline \\[-3pt]
{\bigl \lceil} & 1 & np & 0 & 0 & 1 & {\bigl \rceil}\\
\thvrule & 0 & 1 & 0 & 0 & -p/Q & \hspace{0.5pt} \thvrule\\
\thvrule & 0 & 0 & p & (n\!-\!1) p & p/Q & \hspace{0.5pt} \thvrule\\
\thvrule & 0 & 0 & p & (n\!-\!1) q & 1+p^2/Q & \hspace{0.5pt} \thvrule\\
{\bigl \lfloor} & 0 & 0 & q & p\!+\!(n\!-\!2) q & p^2/Q & {\bigl \rfloor}\\
\\[-3pt]
\hline \\[-3pt]
{\bigl \lceil} & 1 & np & 0 & 0 & 1 & {\bigl \rceil}\\
\thvrule & 0 & 1 & 0 & 0 & -p/Q & \hspace{0.5pt} \thvrule\\
\thvrule & 0 & 0 & 1 & (n\!-\!1) & 1/Q & \hspace{0.5pt} \thvrule\\
\thvrule & 0 & 0 & 0 & (n\!-\!1) (q-p) & 1+p^2/Q-p/Q & \hspace{0.5pt} \thvrule\\
{\bigl \lfloor} & 0 & 0 & 0 & p\!-\!q & p^2/Q-q/Q & {\bigl \rfloor}\\
\\[-3pt]
\hline \\[-3pt]
{\bigl \lceil} & 1 & np & 0 & 0 & 1 & {\bigl \rceil}\\
\thvrule & 0 & 1 & 0 & 0 & -p/Q &\hspace{0.5pt}\thvrule\\
\thvrule & 0 & 0 & 1 & (n\!-\!1) & 1/Q & \hspace{0.5pt}\thvrule\\
\thvrule & 0 & 0 & 0 & 1 & (1+p^2/Q-p/Q)/((n\!-\!1) (q-p)) & \hspace{0.5pt}\thvrule\\
{\bigl \lfloor} & 0 & 0 & 0 & 1 & ((p^2-q)/(p\!-\!q))/Q & {\bigl \rfloor}\\
\\[-3pt]
\hline \\[-3pt]
{\bigl \lceil} & 1 & 0 & 0 & 0 & 1+np^2/Q & {\bigl \rceil}\\
\thvrule & 0 & 1 & 0 & 0 & -p/Q & \hspace{0.5pt} \thvrule\\
\thvrule & 0 & 0 & 1 & 0 & (1-(n\!-\!1)(p^2-q)/(p-q))/Q & \hspace{0.5pt} \thvrule\\
\thvrule & 0 & 0 & 0 & 1 & ((p^2-q)/(p\!-\!q))/Q & \hspace{0.5pt} \thvrule\\
{\bigl \lfloor} & 0 & 0 & 0 & 0 & 0 & {\bigl \rfloor}
\end{array}
\end{equation*}

\section{Statistical properties of the rounding error}
\label{sec:ycor}
In the derivation of square error matrix (\ref{eq:sem1}) in section \ref{sec:error_calc2n} we have assumed the condition of the \textit{Gau\ss-Markov} theorem which is that the error $\varepsilon$ in the weighing can be considered as a random noise, 
is uncorrelated between the $N$ different measurements $y_j$, and has zero mean. 
In this section we investigate the statistical properties of a rounding error in dependence of the resolution of the used measurement scale and the variance in the ground truth weights.
As a result, we obtain an analytic expression for the measurement variance and a condition under which the \textit{Gau\ss-Markov} theorem holds for a rounding error.
\subsection{Analytical considerations}
We recall formula (\ref{eq:linModel}), respectively 
(\ref{eq:linEqSys}), for the linear measurements model which states that the measured weight $y$ corresponds to the real weight $w$ plus a constant offset $m_0$ and some error $\varepsilon$ which is due to the limited precision of the measurement scale, i.e.
\begin{equation}
\varepsilon  = y - \mathbf{m x}\,,
\end{equation}
As in section \ref{sec:Simulations}, we refer in the following to the error-free measurement 
\begin{equation}
    y_{GT} := \mathbf{m x} = w + m_0
\end{equation}
as ground truth measurement. 

In case the scale precision is $a$, the weighing error can be considered as the remainder of rounding $y_{GT}$ to multiples of $a$ (see also \eqref{eq:rounding}), i.e. 
\begin{eqnarray}
r(y_{GT}) =  a\,[y_{GT}/a] - y_{GT}
\end{eqnarray}
where $[ \cdot ]$ denotes the nearest integer number. This implies that the error is a periodic function of $y_{GT}$ with values limited to the interval $]\!-\!a/2\,,a/2\,]$.

Now we consider the case the real weights $w$ vary randomly around a mean value $\mu_w$ with a variance $\sigma_w^2$ described by a general probability distribution $\varphi(w\, |\,\mu_w,\sigma_w^2)$ as, e.g., a Normal distribution.
Then the ground truth measurements $y_{GT}$ will show the same variance $\sigma_w^2$ around a shifted mean value 
   $\mu_y=\mu_w+m_0,$
described by the same general probability distribution
$\varphi(y_{GT}\, |\,\mu_y,\sigma_w^2)$. 
Without loss of generality we can express the general probability distribution based on a distribution $\varphi_0$ with mean $0$ and variance $1$ as
\begin{equation}
\label{eq:yGTDistr}
\varphi(y_{GT}\,|\,\mu_y,\sigma_w^2)
= \frac{1}{\sigma_w}
\varphi_0\left(\frac{y_{GT}\!-\mu_y}{\sigma_w}\right)\,.
\end{equation}
In order to compute the expectation value of the error $\varepsilon$ or any function of it, we need the probability distribution of $\phi(\varepsilon)$ of the error over the interval $]\!-\!a/2\,,a/2\,]$, which can be obtained from the probability distribution $\varphi(y_{GT})$ as
\begin{equation}
\label{eq:DeltayDistrDef}
\phi(\varepsilon) = \int_{-\infty}^{+\infty} \delta(r(y_{GT}\!-\!\varepsilon)) \, \varphi(y_{GT}\, |\,\mu_y,\sigma_w^2)\, {\rm d}y_{GT}\,
\end{equation}
with $\delta$ being the delta-distribution. 
As $\varepsilon\!=\!r(y_{GT})$ for $y_{GT}\!=\!\varepsilon$ modulo integer multiples of the reading precision $a$, the integral (\ref{eq:DeltayDistrDef}) becomes an infinite sum 
\begin{equation}
\begin{aligned}
\label{eq:DeltayDistrGen}
\phi(\varepsilon\,|\,\mu_y,\sigma_w^2) &= \sum_{k=-\infty}^{+\infty} 
\varphi(\varepsilon\!+\!ka\, |\,\mu_y,\sigma_w^2)\,\\
{}&= \frac{1}{\sigma_w}\sum_{k=-\infty}^{+\infty} 
\varphi_0\left(\frac{\varepsilon\!-\mu_y\!+\!ka}{\sigma_w}\right)\,.
\end{aligned}
\end{equation}
This sum can be considered as the periodic probability function
$\phi$ of the error, generated from the probability distribution $\varphi$ of the ground truth measurement by "wrapping it up" on the cylinder with circumference $a$.\\

By definition distribution (\ref{eq:DeltayDistrGen}) is periodically depending on the mean $\mu_y$ of the ground truth. 
If the ratio $a/\sigma_w$ between the scale resolution and the weight variance is sufficiently small, the sum in (\ref{eq:DeltayDistrGen}) can be approximated by an integral
\begin{equation}
\begin{aligned}
\label{eq:DeltayDistrInt}
\phi(\varepsilon\,|\,\mu_y,\sigma_w^2) &= 
\frac{1}{a}\sum_{k=-\infty}^{+\infty} 
\varphi_0\left(\frac{\varepsilon\!-\mu_y}{\sigma_w}+k\frac{a}{\sigma_w}\right)\,\frac{a}{\sigma_w}\\
&\approx \frac{1}{a}\int_{-\infty}^{+\infty} 
\varphi_0\left(\frac{\varepsilon\!-\mu_y}{\sigma_w}+t\right)\,{\rm d}t\,= \frac{1}{a},
\end{aligned}
\end{equation}
thus becoming constant. 
This means that in this limit the rounding error follows a uniform distribution on the interval $]\!-\!a/2\,,a/2\,]$ as has therefore an expectation value and variance of 
\begin{equation}
\label{eq:exp_equal}
\langle\varepsilon\rangle=0,\qquad
\langle\varepsilon^2\rangle=a^2/12
\end{equation}
as one can easily compute.

%
\subsection{Simulations}
\begin{figure}[h]
\includegraphics*[width=0.49\textwidth]{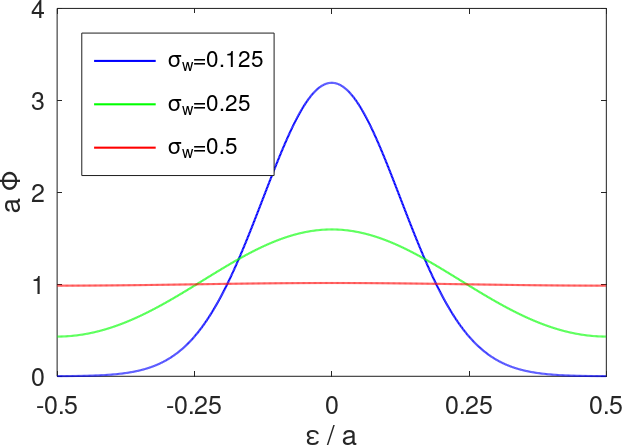}
\hspace*{0mm}
\includegraphics*[width=0.49\textwidth]{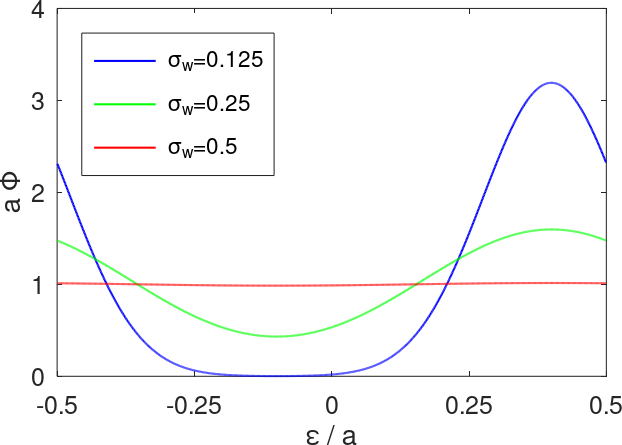}
\caption{Probability distribution of the rounding error assuming an underlying normal distribution with variance $\sigma_w^2$ and mean $\mu_y=0$ (left), respectively $\mu_y=0.4$ (right) and of the ground truth weights. 
\label{fig:periodicGauss}}
\end{figure}

Figure \ref{fig:periodicGauss} shows examples of the probability distribution of the rounding error $\varepsilon$ assuming an underlying Gaussian distribution of the weights.
One finds that for a standard deviation of the weights smaller than half the resolution, i.e. $\sigma_w\!<\!a/2$, the shape of the error distribution resembles that of the underlying Gaussian centered around the measurement error of the mean $\mu_y$. 
This distribution continues periodically across the boarders of the interval $]-a/2,a/2\,]$, as can be recognized in the right figure, where the center of the distribution is close to the right boarder.
If the standard deviation of the weights is increased to half the resolution or more, the periodic wrapping of the Gaussian leads to an entire leveling of the distribution becoming thus constant, as predicted above in formula (\ref{eq:DeltayDistrInt}), i.e.
\begin{equation}
\label{eq:sigmaa2}
    \phi(\varepsilon\,|\,\mu_y,\sigma_w^2) = \frac{1}{a}\,\qquad \mathrm{for} \quad \sigma_w\!>\!a/2.
\end{equation}

\begin{figure}[h]
\includegraphics*[width=0.49\textwidth]{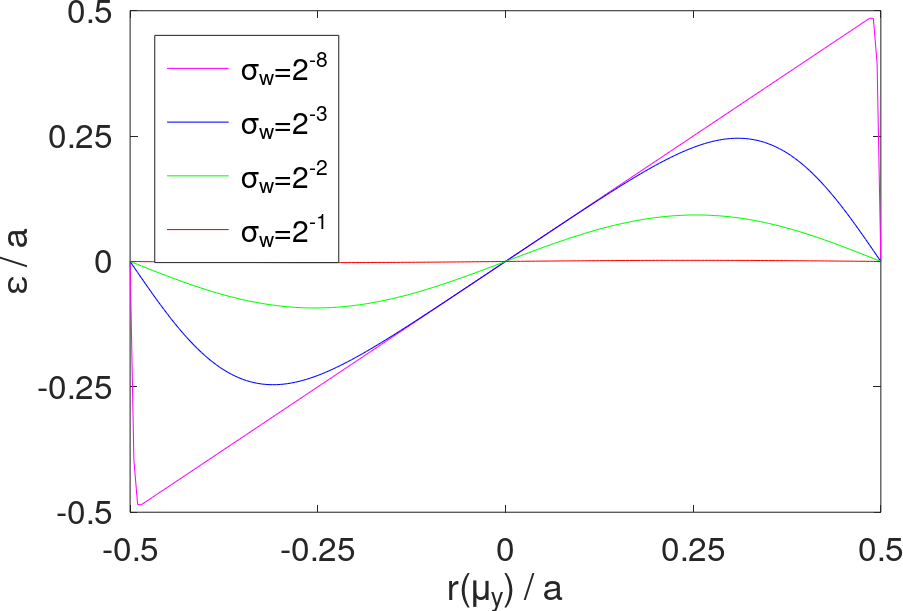}
\hspace*{0mm}
\includegraphics*[width=0.49\textwidth]{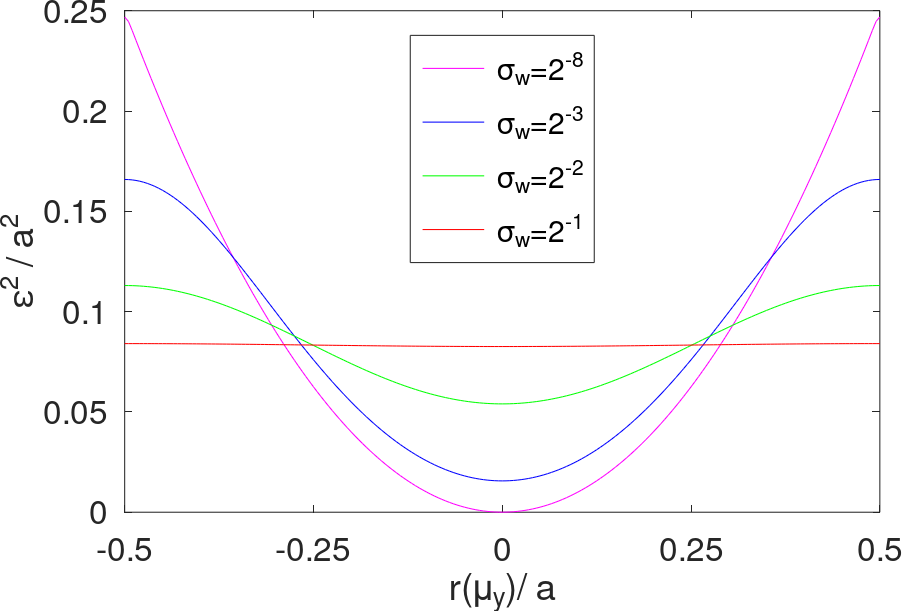}
\caption{Expectation value of the errors $\langle \varepsilon \rangle$ (left) and expectation value of the square error $\langle \varepsilon^2 \rangle$ (right) as a function of the remainder $r(\mu_y)$ of the mean ground truth measurement.}
\label{fig:Edy}
\end{figure}

Computing the expectation value of the error using probability distribution (\ref{eq:DeltayDistrGen}), 
one obtains the results shown in the left figure (\ref{fig:Edy}).
The error expectation value $\langle\varepsilon\rangle$ is plotted as a function of the ground truth mean measurement $\mu_y$ for different values 
of the ground truth weight variance $\sigma_w^2$.
One recognizes that for very small weight variance $\sigma_w\!\ll\!a/2$,
the expectation value of the error corresponds to the remainder $r(\mu_y)$
of the mean ground truth.  For large variance, the leveling of the distribution causes the error expectation value to go to zero, which is the result of the uniform probability distribution (\ref{eq:DeltayDistrInt}) and thus the case if $\sigma_w\!>\!a/2$.\\

Figure \ref{fig:Edy} shows the mean measurement error (left) as well as the mean square error (right) for different weight variance. One recognizes 
for small variance a strong dependence on the position of the mean 
$\mu_y$. In the limit $\sigma_w\!\ll\!a/2$ we find the square error being a parabola with $0$ at the center and $(a/2)^2$ at the boarder of the interval $[-a/2,a/2]$.
On the contrary for $\sigma_w\!>\!a/2$ the expectation value of the square error becomes constant with a value of  $\langle\varepsilon^2\rangle=a^2/12$ as analytically derived  above (see \eqref{eq:exp_equal})

\subsection{Conclusions}
In summary, we have demonstrated that, if the measured weights follow a Gaussian distribution with a variance $\sigma_w$ larger than half of the scale resolution $a$, the rounding error will be equally distributed.
As a result, the rounding error will be mean value free, being a condition of the \textit{Gau\ss-Markov} theorem, and the error variance is 
\begin{equation}
\label{eq:sigma_a}
\sigma^2 = \text{var}(\varepsilon)=a^2/12.
\end{equation}
Inserting this relation \eqref{eq:sigma_a} into formula \eqref{eq:sem2} and the more specific formulas \eqref{eq:delta_m}, one obtains an explicit analytic expression for the regression error in the estimated masses.

To apply the above result to our combined weighing scheme, we have to consider that weights are combinations of several masses. 
However, if the $n$ unknown masses follow a Gaussian distribution with $\sigma_m\!>\!a/2$, also the combination of $k$ masses will fulfill this condition, as combining several masses increases the variance of the weight distribution. 

This fact gives rise to the hope that the limits of the \textit{Gau\ss-Markov} theorem can be pushed further towards smaller masses by combining them in the weighing scheme.
The analysis of the statistical distribution of combined masses requires, however, further investigations. 
One aspect is that a fixed set of $n$ masses leads to correlations in the total weight which limits their variations. 
Moreover, combining different numbers $k$ of masses does not result in a uni-modal weight distribution, as a Gaussian, but in a multi-modal distribution. 
We therefore leave this further statistical investigation to future work.

For the time being we remain with the conclusion that if the $n$ unknown masses follow a Gaussian distribution with $\sigma_m > a/2$, the conditions of the \textit{Gau\ss-Markov} theorem is fulfilled for the combined weighing schemes we proposed in this work. 

Looking on the set-up described in section \ref{ssec:setup} we find that the variance of the used weights is approximately half of the scale accuracy of $a = \SI{20}{\g}$ such that we can assume that for our experiments the \textit{Gau\ss-Markov} condition was fulfilled.


\section*{Acknowledgments}
We are grateful to Gabriele Backes, Katharina Boguslawski, Una Karahasanovic, Hans-Jürgen Korsch, and Valeria Serchi for critically reading the manuscript and giving many useful recommendations.
\clearpage
\bibliographystyle{unsrt}


\end{document}